\documentclass[10pt,a4paper,twocolumn,floatfix,superscriptaddress]{revtex4}
\usepackage{amsmath}
\usepackage{amsfonts}
\usepackage{amssymb}
\usepackage{graphicx}
\usepackage{array,tabularx}
\usepackage{color}
\def\rl{\phi_{rlp}}
\def\rc{\phi_{rcp}}
\def\vrl{\phi_{rvlp}}

\begin{document}
\title{Random very loose packs}
\author{Massimo Pica Ciamarra}
\email{picaciam@na.infn.it}
\affiliation{CNISM and Department of Information Engineering, Second University of Naples, 81031 Aversa (CE), Italy}
\affiliation{Dip. di Scienze Fisiche, Universit\'a di Napoli Federico II, 80129, Naples, Italy}
\author{Antonio Coniglio}
\affiliation{Dip. di Scienze Fisiche, Universit\'a di Napoli Federico II, 80129, Naples, Italy}
\affiliation{INFM-CNR, Coherentia, Napoli, Italy}

\begin{abstract}
We measure the number $\Omega(\phi)$ of mechanically stable states of volume fraction $\phi$ of a granular assembly under gravity.
The granular entropy $S(\phi) = \log \Omega(\phi)$ vanishes both at high density, at $\phi \simeq \rc$, and a low density, at $\phi \simeq \vrl$, where $\vrl$ is a new lower bound we call {\it random very loose pack}. $\rl$ is the volume fraction where the entropy is maximal. These findings allow for a clear explanation of compaction experiments, and provide the first first-principle definition of the random loose volume fraction. In the context of the statistical mechanics approach to static granular materials, states with $\phi < \rl$ are characterized by a negative temperature.
\end{abstract}
\maketitle

If you pour grains in a container, you will find their density to be always enclosed between an upper and an lower bound, known as the random close $\rc$ and the random loose $\rl$ volume fractions. While $\rc$ has been deeply investigated~\cite{packs}, $\rl$ has received considerable less attention, and at the present time it has not a precise definition, {\it except for the implication that it represents the loosest possible, random packing that is mechanically stable}~\cite{onoda}.
Here we show that $\rl$ is actually the loosest possible random packing which is mechanically stable {\it one can achieve by pouring grains}, and that there are actually many mechanically stable states of volume fraction $\phi < \rl$. We find that the entropy $S(\phi) = \log \Omega(\phi)$, where $\Omega(\phi)$ is the number of disordered mechanically stable states of volume fraction $\phi$, goes to zero at $\rc$ and at $\vrl < \rl$, where $\vrl$ is a new lower bound we call {\it random very loose volume fraction}. $\rl$ turns out to be the volume fraction where the entropy is maximal. We give an interpretation of both $\rc$, $\rl$ and $\vrl$ in the context of the statistical mechanics of dense granular materials.

\begin{figure}
\begin{center}
\includegraphics[scale=0.95]{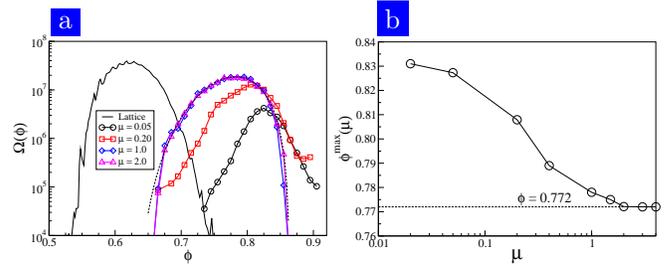}
\end{center}
\caption{\label{fig:entropy} 
{\bf a} Number of mechanically stable configurations of $N = 20$ disks on a lattice, and on the continuous for different values of the interparticle friction coefficient. {\bf b} The volume fraction where $\Omega_\mu(\phi)$ is maximal approaches the estimated value of $\rl$ as the friction coefficient increases.}
\end{figure}

To measure the granular entropy, we have considered a system of $N = 20$ disks of diameter $D$ under gravity, placed in a box with horizontal dimension $l_x = 4D$, where we use periodic boundary conditions. First, we have considered a coarse grained lattice model of the system,
as described in the supplementary materials available on-line~\cite{supporting}.  The lattice model correspond to a discretization of the phase space of the system in cells of size $\delta l ^{2N}$; as we use $\delta l = D/5$, the number of cells is approximately $(l_x/\delta l)^{2N} \simeq 20^{40}$.  Despite this huge number,
the entropy of the lattice model can be computed exactly, as it is possible to reduce the exploration of the phase space to only those cells
that correspond to mechanically stables state which, for the system we are considering, are slightly more than $10^9$.  The number of stable configurations at each volume fraction $\phi$, $\Omega_{L}(\phi)$, is shown in Fig.~\ref{fig:entropy}a. As the lattice states are not actually mechanically stables on the continuous, $\phi$ spans unusual small values. 
In principle, the entropy of the system on the continuous could be determined investigating that of the lattice model as $\delta l$ decreases; yet, as the number of configurations increases as $\delta l^{-2N}$, it is clear that this approach cannot be carried out.

\begin{figure}
\begin{center}
\includegraphics[scale=0.85]{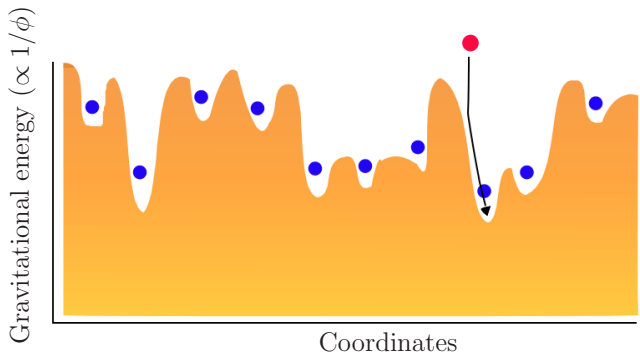}
\end{center}
\caption{\label{fig:landscape}
Schematic landscape of the energy as a function of the configurational coordinates. The red disk represents a random configuration, used as starting point of energy minimization procedures to find mechanically stable states (energy minima). The blue disks represent the `almost stable' initial configurations we use as starting points of our minimization procedure (see text).
}
\end{figure}

Even though the configurations stable on the lattice are not stable on the continuous, they provide a sampling of the phase space
of the mechanically stable states of the continuous system. In fact, as schematically depicted in Fig.~\ref{fig:landscape}, each configuration stable on the lattice is close to energy minima, which can be reached using the lattice generated configuration as starting point of Molecular Dynamics (MD) of sedimentation under gravity. We now make an ansatz, which will be justified {\it a posteriori}, and assume that the lattice generated configurations sample uniformly the phase space of the continuous system. 
Under this assumption, it is possible to determine the entropy of the continuous system (up to a constant factor) by performing the MD simulations of sedimentation, and counting how many minima occurs in the volume fraction range $\phi$, $\phi + \delta\phi$.
%Under the above assumption, the entropy of the system on the continuous can be derived from the entropy of the lattice model,
%as the configurations which are stable on the lattice can be considered as a uniform sampling of all of the minima. A configuration stable on the lattice, in fact, sample an energy minima because, even thought it is not actually mechanically stable, it is close to an energy minima as schematically depicted in Fig.~\ref{fig:landscape}. The sampled minima can be determined using the lattice stable configurations as initial configurations of Molecular Dynamics (MD) simulations of sedimentation under gravity. 
We have performed these simulations using a standard model for the grain-grain interaction~\cite{Silbert01}, and considering different values of the friction coefficient $\mu$.  The difference between this approach to explore the phase space of the mechanically stable states and previous ones~\cite{nowak,bideau,Matthias,PicaCiamarra06} in only in the choice of the initial conditions, as illustrated in Fig.~\ref{fig:landscape}.
%We now describe how it is possible to estimate the entropy of the continuous system.
%The idea is that, while the configurations which are stable on the lattice are not stable on the continuous, they can be considered as `almost stable', meaning that they are close to an energy minimum (a mechanically stable state) of the continuous system.
%In order to probe these minima, one can therefore use the lattice generated configurations as starting point of Molecular Dynamics simulations of sedimentation under gravity. We have performed these simulation using a standard model for the grain-grain interaction~\cite{Silbert01}, and considering different values of the friction coefficient $\mu$. We stress that the difference between this approach to sample the phase space and previous ones~\cite{nowak,bideau,Matthias,PicaCiamarra06}, is only in the choice of the initial conditions as schematically depicted in Fig.~\ref{fig:landscape}.

The number $\Omega_\mu(\phi)$ of mechanically stable configurations of volume fraction $\phi$, when the static friction coefficient between the grains is $\mu$, is therefore estimated as:
\begin{eqnarray}
\Omega_\mu(\phi) &=& \int d\phi' \Omega_{L}(\phi')P_\mu(\phi' \to \phi) - \\
 &-&\frac{1}{2}\int d\phi' \int d\phi'' \Omega_{L}(\phi')\Omega_{L}(\phi'')P_\mu(\phi', \phi'' \to \phi). \nonumber
\end{eqnarray} 
Here the first term accounts for the probability that the sedimentation procedure transforms a lattice-stable configuration
of volume fraction $\phi'$ in a stable configuration of volume fraction $\phi$, the static friction coefficient being $\mu$. The second term takes into account the possibility that two different lattice-stable configurations actually lie in the same energy minimum, and assures that we don't count an energy minima more than once. $P_\mu(\phi', \phi'' \to \phi)$ is the probability that an almost stable configuration of volume fraction $\phi'$, and one of volume fraction $\phi''$, give rise to the same mechanically stable state of volume fraction $\phi$. 
By performing simulations of sedimentation using as initial states some of the lattice generate configurations, we have estimated $P_\mu(\phi' \to \phi)$ and $P_\mu(\phi', \phi'' \to \phi)$~\footnote{To estimate $P_\mu(\phi', \phi'' \to \phi)$ we have considered two configurations to be different if there is a disk in position $x,y$ in one of the configurations, and there is not a disk in position $x',y'$ in the other configuration, with 
$(x-x')^2+(y-y')^2 < \delta^2$, and $\delta = D/5$. Increasing $\delta$ as the effect of multiplying $\Omega(\phi)$ by a constant factor ($ < 1$), as more states are considered to be equal, and does not change neither the position of the maximum of $\Omega(\phi)$, nor the values of $\phi$ where $\Omega(\phi)$ vanishes.}. 
Precisely, we have performed a total of $2\;10^4$ simulations for each value of $\mu$, each simulation ending when the kinetic energy becomes zero within numerical errors; when this is the case, the system is always found in an energy minimum as all of the eigenvalues of the Hessian are negative, and therefore no farther relaxation occurs.
Figure~\ref{fig:entropy}a shows $\Omega_\mu(\phi)$ for $\mu = 0.05,0.2,1.0$ and $2.0$. 
For all values of $\mu$, the entropy $S(\phi) = \log \Omega(\phi)$ has a maximum, and decreases smoothly for smaller and larger values of $\phi$. At high $\phi$, $\Omega(\phi)$ does not vanishes at $\rc \simeq 0.82$ as we observed the formation of crystalline structures, as usual in 2D dimensional systems. The entropy should be an increasing fraction of $\mu$, and we observe that this is actually the case but at very high density, where the limit of the coarse graining procedure used in the lattice model shows up~\footnote{Results obtained via the study of smaller systems, where a 
more refined coarse grained model could be used, shows that the volume fraction range where the entropy is non monotonously increasing with $\mu$ shrinks as the ratio $D/l$ between the particle diameter and the lattice spacing of the coarse grained model increases.}.

The value of the volume fraction $\phi^{\rm max}$ where the entropy is maximal depends on the friction coefficient $\mu$. A plot of $\phi^{\rm max}$ as a function of $\mu$, as shown in Fig.~\ref{fig:entropy}b, reveals that, at high friction, $\phi^{\rm max}$ approaches the limiting value $0.772$. This is a common estimate of $\rl$ for frictional systems~\cite{dong06,Hinrichsen}. Accordingly, we suggest that $\rl$ can be defined as the value $\phi^{\rm max}$ where the entropy is maximal. This definition reproduces the $\mu$ dependence of $\rl$, previously suggested~\cite{Texas}, and does also explain compaction experiment as discussed below. These evidences clarify that there are many disordered mechanically stable states of volume fraction $\phi < \rl$, and suggest to introduce a lower bound for the volume fraction of a mechanically stable disordered assembly of particles, we term {\it random very loose volume fraction}, whose estimation in 2D is $\vrl \simeq 0.675$. 
To stress that the presence of mechanically stable states with very low volume fraction is not a finite-size effect, we have also devised a procedure to construct low density states with an arbitrary number of disks, described in the supplementary information~\cite{supporting}; a web-applet based on this procedure is also available~\footnote{http://smcs.na.infn.it/rvlp/rvlp.html}.
\begin{figure*}[t!]
\begin{center}
\includegraphics*[scale=1]{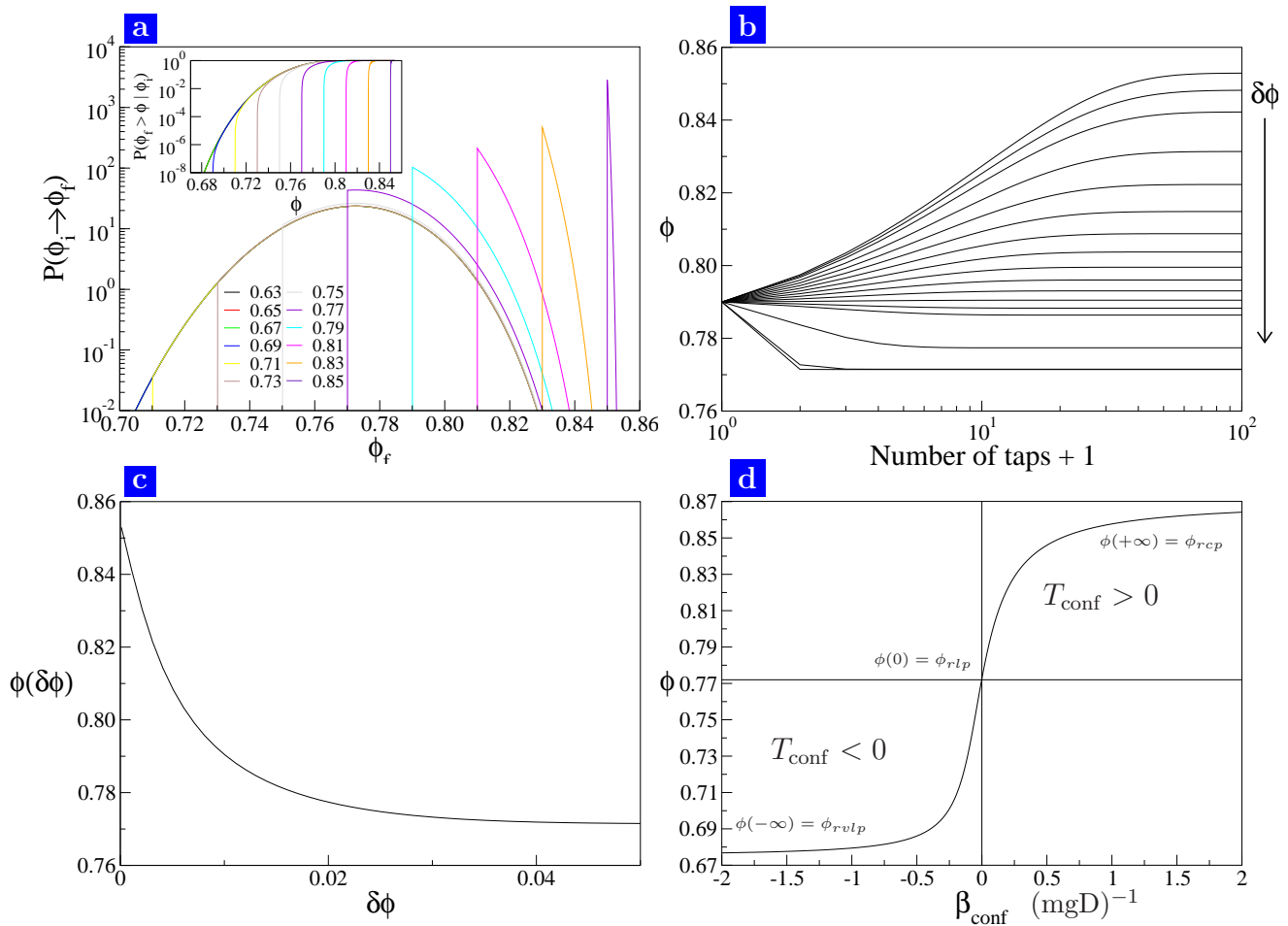}
\end{center}
\caption{\label{fig:varie}
{\bf a} The main panel (inset) shows the probability $P(\phi_{i} \to \phi_f)$ ($P(\phi_{f} > \phi|\phi_i)$) that a sedimentation procedure which starts from an unstable state configuration of volume fraction $\phi_i$ ends in a stable state of volume fraction $\phi_f$ (greater than $\phi$). {\bf b} Evolution of the volume fraction in a sequences of taps. {\bf c} The volume fraction reached at stationary in a tapping dynamics as a function of the intensity of the tap $\delta\phi$. {\bf d} Equation of state in the statistical mechanics approach to granular media.}
\end{figure*}

Our results for the entropy are based on the assumption that the configurations stable on the lattice provide a uniform sampling of the minima of the continuous system. While this assumption could be in principle validated by investigating lattice models with a small value of $\delta l$, we note here that the assumption works, at least as a first approximation, because it allows to reproduce all expected features of the entropy, as well as to explain other experimental evidences found both in 2 dimensions and in 3 dimensions. In fact, the shape of the entropy we have determined 1) decreases as $\phi$ varies between $\rl$ and $\rc$ as recently found~\cite{Aste,Makse2}, 2) increases with the friction coefficient as expected, 3) explains the dependence of $\rl$ on the friction coefficient, 4) explains compaction experiments~\cite{Texas} and why states with volume fraction smaller than $\rl$ have not been experimentally observed (see below). 
Moreover, the non-monotonic behavior of the entropy shown in Fig.~\ref{fig:entropy}a is expected from the simple consideration that the number of mechanically stable configurations of our system can be expressed as $\Omega(\phi) = \Omega_{gas}(\phi)\Pi(\phi)$, where $\Omega_{gas}(\phi)$ is the number of configurations of volume fraction $\phi$ {\it without} taking into account the stability criterion, and $\Pi(\phi)$ is the fraction of configurations of volume fraction $\phi$ which are stable. As $\phi$ decreases, $\Omega_{gas}(\phi)~\propto \exp(1/\phi)$ exponentially increases because there is more space where to put the disks, while $\Pi(\phi)$ decreases. Therefore their product, as well as the entropy $S(\phi)$, must have a maximum as we have found.

We now show that the identification of $\rl$ with the volume fraction where the entropy is maximal allows to explain
compaction experiments~\cite{Matthias,PicaCiamarra06}, as well as why states with $\phi < \rl$ have not been previously observed. 
To this end, we first consider what is the probability distribution $P(\phi_i \to \phi_f)$ that a random unstable state of volume fraction $\phi_i$ becomes a mechanically stable state of volume fraction $\phi_f$ in a sedimentation procedure. 
Neglecting any dynamical effect, an assumption which is reasonable when the sedimentation procedure is very slow, 
%as when the system is immersed in a fluid and driven by flow pulses as in\cite{Matthias,Texas},
we assume that all mechanically stable states with volume fraction $\phi_f  \leq \phi_i$ have the same probability to be found via the sedimentation procedure. Accordingly,  $P(\phi_{i} \to \phi_f)$ is simply proportional to $\Omega(\phi_f)$:
\begin{equation}
 P(\phi_{i} \to \phi_f) = \frac{\Omega(\phi_f)}{\int_{\phi_i}^{\rc}\Omega(\phi')d\phi'} \;\;\ \textnormal{if } \phi_i \leq \phi_f
\label{eq:compaction}
\end{equation}
while $P(\phi_{i} \to \phi_f) = 0 $ if $\phi_i > \phi_{f}$, as during a sedimentation process the density increases.

To calculate this and other quantities we approximate in the following the granular entropy with
$S(\phi) = \log \Omega(\phi) \propto (\phi-\vrl)^\alpha(\rc-\phi)^\beta$ (see dashed line in Fig.~\ref{fig:entropy}a), but different approximations give rise to very similar results. We show in Fig.~\ref{fig:varie}a, for different values of $\phi_i$, both the probability distribution $P(\phi_i \to \phi_f)$ and the probability distribution that $\phi_f  > \phi$, $P(\phi_f > \phi\;|\;\phi_i)$ (inset). 
If $\phi_i < \rl$, $P(\phi_i \to \phi_f)$ has a maximum at $\phi_f \simeq \rl$. If $\phi_i > \rl$, on the contrary, the maximum of $P(\phi_i \to \phi_f)$ is at $\phi_f \simeq \phi_i$. 
These results indicate that in usual sedimentation procedures packs of volume fraction smaller that $\rl$ are not observed just because of entropic reasons: the number of states with  $\phi < \rl$ is exponentially too small with respect to the number of packs of volume fraction $\phi \simeq \rl$.

Eq.~\ref{eq:compaction} could be used to describe compaction experiments, where a granular system is subject to a sequence of taps. In a single tap, a mechanically stable state of volume fraction $\phi_n$ first becomes an unstable state of volume fraction $\phi_n-\delta\phi$, due to the mechanical energy which is given to the system, and then settles down in a new mechanically stable state of volume fraction $\phi_{n+1}$. $\delta\phi > 0$ is related to intensity of the tap: the stronger the tap, the larger $\delta\phi$. After many taps it is observed that the volume fraction reaches a steady state $\phi(\delta\phi)$ which depends on the intensity of the tap~\cite{nowak,bideau,Matthias,PicaCiamarra06}. Fig.~\ref{fig:varie}b shows that this kind of behavior is reproduced by considering $\phi_n-\delta\phi = \phi_i$ and $\phi_{n+1}(\phi_n,\delta\phi) = \int \phi_f P( (\phi_n-\delta\phi) \to \phi_f) d\phi_f$. As the number of taps increases the volume fraction reaches a steady value $\phi(\delta \phi)$, which is the fixed point of the recurrence relation $\phi_{n+1} = \phi_{n+1}(\phi_n,\delta\phi)$. The dependence of $\phi(\delta \phi)$ on $\delta\phi$, which is shown in Fig.~\ref{fig:varie}c, reproduces the dependence of the volume fraction reached at stationary on the intensity of vibration in compaction experiments~\cite{nowak,bideau,Matthias,PicaCiamarra06}.
The identification of $\rl$ with the volume fraction where the entropy is maximal is therefore consistent with previous works
which have considered $\rl$ as being the loosest pack achieved via sedimentation or tapping procedures~\cite{Matthias,Texas,onoda}.

The results presented here have an interesting interpretation in the context of the statistical mechanics approach to static granular media~\cite{edwards89}. In this theoretical framework, one considers a granular system subject to a dynamics allowing for the exploration of the mechanically stable states, and assumes that all states of given volume $V$ are visited with the same probability. For hard spheres under gravity the energy $E$ is dominated by the gravitational energy, and therefore $V \propto E  \propto 1/\phi$. This hypothesis has been proved numerically at $\phi \simeq \rc$ for a system under shearing~\cite{Makse}, and in the density range $\rl < \phi < \rc$ for particles driven by flow pulses~\cite{PicaCiamarra06}. Accordingly, one can introduce a granular temperature, known as configurational temperature or compactivity, defined as $T_{conf}^{-1} = \beta_{conf} = \partial S /\partial E$. The dependence of $\phi$ on $\beta_{conf}$ (see~\cite{supporting} for calculation details) is shown in Fig.~\ref{fig:varie}d, and clarifies that in this theoretical framework $\rc,\rl$ and $\vrl$ are defined as follows: $\rc = \phi(\beta_{conf} \to +\infty)$, $\rl = \phi(\beta_{conf} = 0)$, and $\vrl = \phi(\beta_{conf} \to -\infty)$. The density range that is not usually observed, with volume fraction enclosed between $\vrl$ and $\rl$, is that where the granular temperature is negative.

%\end{document}

%\documentclass{article}
%\usepackage{amsmath}
%\usepackage{amsfonts}
%\usepackage{amssymb}
%\usepackage{graphicx}
%\oddsidemargin  1cm
%\textwidth 14cm
%\textheight 22cm
%\def\rl{\phi_{rlp}}
%\def\rc{\phi_{rcp}}
%\def\vrl{\phi_{rvlp}}
%\onecolumn
\pagestyle{empty}
\begin{widetext}
\newpage
%\begin{document}
\noindent {\Large \bf Supplementary information, \\ Methods}\\
{Hard disks on a lattice}\\
\\
\noindent In order to uniformly sample the phase space of the mechanically stable configurations of an hard disk system under gravity, we have used as starting point of a sedimentation procedure configurations which are `almost stable', meaning that they are close to an energy minima. We have generated these configurations using a lattice model obtained by coarse graining an hard disk system, the `almost stable' configurations being defined as those that are stable according to the stability rules defined on the lattice. \\
\\ \noindent In the model, the presence of a grain in a given lattice site hinders the occupancy of nearby lattice sites due to steric constraints. For instance, the grain shown in Fig.~\ref{fig:model} forbids the occupancy of the lattice sites marked with a star (as well as of all the lattice sites covered by the disk). As we work under gravity, we consider a grain to be stable if it is supported by both a grain on its left and by a grain on its right, i.e. if at least one of the positions marked with $L$ and one of the positions marked with $R$ in Fig.~\ref{fig:model} are occupied. We also consider the possibility that two grains $A$ and $B$ cooperate to their stability. Precisely, if grain $A$ is in position ($i,j$) and grain $B$ in position ($i+D,j)$, and both one of the $L$-positions of grain $A$ and one of the $R$-positions of grain $B$ are occupied, then both grains are mechanically stable. Finally, a grain resting on the bottom or on the first layer ($j = 0,1$) of the lattice is also considered as mechanically stable. A configuration is mechanically stable if all of its grains are mechanically stable. 
\begin{figure}[!ht]
\begin{center}
\includegraphics*[scale=0.3]{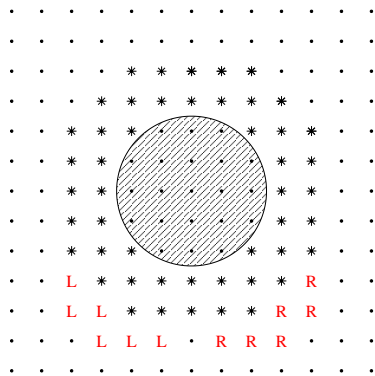}
\end{center}
\caption{\label{fig:model} Disks on a lattice. If a disk is present in the system as in figure, then all of the lattice sites covered by the disk as well as the lattice sites marked with $\ast$ cannot be occupied by other disks due to steric constraints. The positions marked with $L$ and $R$ are used to define the mechanically stable configurations in the text.
}
\end{figure}
\noindent The stability rules implies that, if a grain at a given heigh $h$ is not mechanically stable, then it is not possible to make it stable by adding grains at an height $h' > h$. By exploiting this feature it possible to devise a procedure to selectively generate all of the mechanically stable configurations of this model, without the need to generate the unstable ones.
As we want to evenly sample the phase space, we have considered a system of $N = 20$ disks on a square lattice with horizontal size $4D$ and lattice constant $D/5$, for which it is possible to generate {\bf all} of the stable configurations. 
A typical lattice generated configuration, together with the configuration reached after sedimentation, is shown in Fig.~\ref{fig:configurations}. 

\begin{figure}[!ht]
\begin{center}
\includegraphics*[scale=0.5]{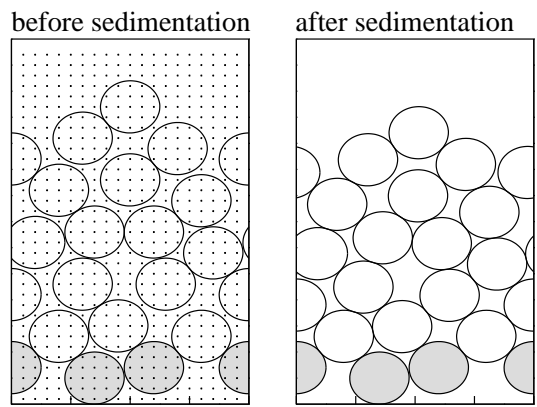}
\end{center}
\caption{\label{fig:configurations} A lattice generated configuration, and the corresponding configuration reached after sedimentation.
}
\end{figure}
It is not possible to investigate significatively larger values of $N$, as the number of mechanically stable configurations grows exponentially with $N$, as shown in Fig.~\ref{fig:number}.
\begin{figure}[!ht]
\begin{center}
\includegraphics*[scale=0.33]{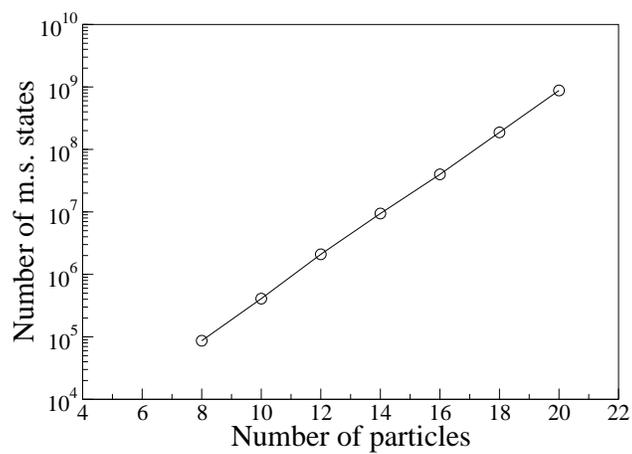}
\end{center}
\caption{\label{fig:number} The number of lattice-stable configurations increases exponentially with the number of grains. 
}
\end{figure}
\newpage
~
\newpage
\noindent {\Large \bf Supplementary information, \\Methods}\\
{\bf Generation of very loose mechanically stable random states}\\
\\
\noindent Via the combined use of the lattice model to generate `almost stable' configurations, and of Molecular Dynamics simulations of sedimentation under gravity, it is possible to uniformly sample the phase space of an hard disks system, and to generate mechanically stable states of with $\phi < \rl$. This approach has the advance of allowing an uniform exploration of the phase space, but is limited to systems with a small number of particles.

Here we present a procedure to generate mechanically stable states with a very small value of $\phi$ with an arbitrary number of particles, in the limit of hard particles with very high friction. A web-applet based on this procedure is availabe (http://smcs.na.infn.it/rvlp/rvlp.html). Figure~\ref{fig:sample} shows a pack generated with this procedure, with volume fraction (calculated in the area delimited by the two horizontal lines) $\phi \simeq 0.746$.
\begin{figure}[h]
\begin{center}
\includegraphics*{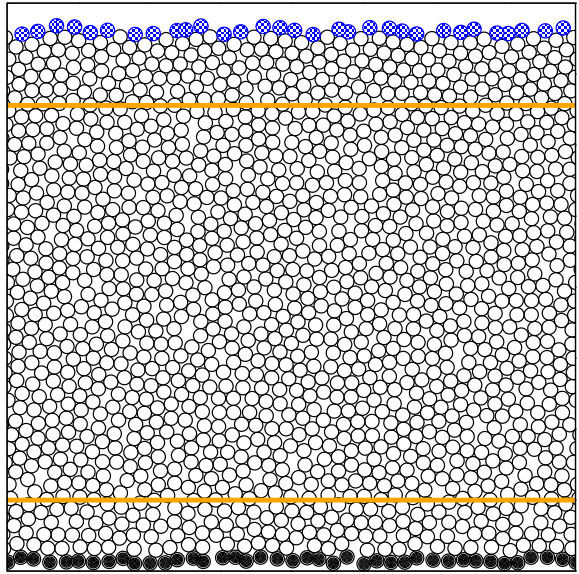}
\end{center}
\caption{\label{fig:sample} A mechanically stable loose with volume fraction $\phi \simeq 0.746 < \rl \simeq 0.772$.}
\end{figure}

The procedure is based on a growing mechanism, where disks of diameter $D$ are added on top of previous placed disks. Given a mechanically stable state, the number of particles can be increased by adding $n$ disks at the same time. If $n=1$, then there is a finite number of positions where the new grain can be placed. These positions can be easily found from geometrical considerations and are show as blue shaded disks in Fig.~\ref{fig:sample}. For $n > 1$, there is an infinite number of mechanically stable configurations the disks $n$ can be arranged. We have considered, for simplicity sake, only $n=2$, and we have placed the two grains according to the following steps:
\begin{enumerate}
\item Choose a disk $A$ on the bed surface and an angle $\theta$, $-\pi < \theta < \pi$.
\item Place a new disk $\alpha$ in contact with disk $A$, the line joining their center making an angle $\theta$ with the positive vertical direction. If the grain overlaps other grains, eliminate the new grain and go back to $1$.
\item Determine the positions where it is possible to add a single grain, the grain being in contact with grain $A$. If there are no positions, go back to $1$.
\item Place a new grain $\beta$ in one of the previously determined positions. The grain will touch a grain $B$. If grain $\alpha$ and $\beta$ are not stable, eliminate them and go back to $1$.
\end{enumerate}

In the limit of hard disks and infinite friction we are considering, the stability of grains $\alpha$ and $\beta$ can be easily determined form geometrical considerations, as illustrated in Fig.~\ref{fig:bridge}. For instace, grain $\alpha$ is stable if $x_A < x_\alpha < x_B$ and $y_\beta > y_P$, where $P$ is the point where the line joining grains $A$ and $\beta$ and the vertical line passing trough the center of grain $\alpha$ intersect. A similar condition holds for grain $\beta$. The resulting bridge is stable if both grain $\alpha$ and grain $\beta$ are stable.

\begin{figure}[h]
\begin{center}
\includegraphics*[scale=0.5]{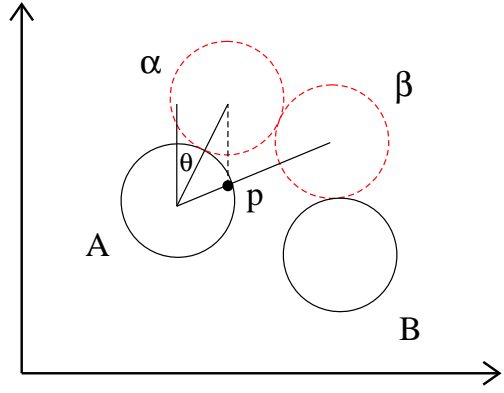}
\end{center}
\caption{\label{fig:bridge} Stability of a bridge.}
\end{figure}

The hypothesys of infinite friction can be relaxed after the packs has been build. In fact, if the pack is used as starting point of soft core Molecular Dynamics simulations of relaxation under gravity, one observes a small compactification, related to the softness of the disks, during which tangential forces build up. Once the pack becomes stable, one can determine the 
the minimum friction coefficient required for the pack to be stable, by measuring the normal $\vec F^{(i)}_n$ and the tangential $\vec F^{(i)}_t$ forces at each contact $i$. The minimum friction coefficient is: $\mu = \max_{i} |\vec F^{(i)}_t|/|\vec F^{(i)}_n|$.
%\end{document}

\newpage
\noindent {\Large \bf Supplementary information, Methods}\\
{\bf The equation of state}\\
\\
\noindent The equation of state shown in Fig.4d, expressing the dependence of $\phi$ on the inverse temperature is easily determined
as 
\begin{equation}
\beta_{conf}(\phi) = \frac{\partial S}{\partial \phi}(\phi)\frac{\partial \phi}{\partial E}(\phi).
\nonumber
\end{equation}
We use for the entropy the expression
\begin{equation}
S(\phi) = A[(\phi-\vrl)^\alpha(\rc-\phi)^\beta],
\nonumber
\end{equation}
which describes the numerical data (see Fig.1a) with $A = 7$, $\vrl = 0.675$, $\rc = 0.87$, $\alpha = 0.28$, $\beta = 0.15$.
The energy $E$ is given by the gravitational energy of the system,
\begin{equation}
E = \frac{1}{2}Nmgh,
\nonumber
\end{equation}
where $N = 20$ is the number of disks, $m$ the mass of each disk, $g$ the gravitational field, and $h$ the height of the system. $h$ is related to the volume fraction $\phi$, as
\begin{equation}
\phi = \frac{N v}{l_x h};~~~~~~~~~~~ v = \pi \left(\frac{D}{2}\right)^2
\nonumber
\end{equation}
where $D$ is the diameter of the disks, and $l_x = 4D$ the horizontal size of our system.\\
From the above expressions, we obtain:\\
\begin{equation}
\frac{\partial S}{\partial \phi}(\phi) = A\left[   \alpha (\phi-\vrl)^{\alpha-1}(\rc-\phi)^\beta - \beta(\phi-\vrl)^\alpha(\rc-\phi)^{\beta-1}      \right]
\nonumber
\end{equation} 
and

\begin{equation}
\frac{\partial \phi}{\partial E}(\phi) = \frac{\partial \phi}{\partial h}\frac{\partial h}{\partial E} = \left(-\frac{Nv}{l_xh^2}\right) \left(\frac{2}{Nmg}\right) = -\frac{2 l_x}{N^2 v mg}\phi^2 = -\frac{16}{N^2\pi}\frac{\phi^2}{mgD}.
\nonumber
\end{equation} 
\end{widetext}
\end{document}